\pdfoutput=1  

\documentclass[11pt]{article}
\usepackage{amsmath}

\usepackage{pdfpages}

\input{tcilatex}
\raggedright

\begin{document}

\date{}
\title{}
\author{}
\maketitle

{\small Risk060809.tex}

\begin{center}
{\LARGE Evaluating Health Risk Models }

\bigskip

Alice S. Whittemore*

Department of Health Research and Policy

Stanford University School of Medicine, 259 Campus Drive, Stanford, CA
94305-5405

alicesw@stanford.edu

Tel: 650-723-5460 Fax: 650-725-6951

\bigskip

{\Large SUMMARY}
\end{center}

\noindent \qquad Interest in targeted disease prevention has stimulated
development of models that assign risks to individuals, using their personal
covariates. We need to evaluate these models, and to quantify the gains
achieved by expanding a model with additional covariates. \ We describe
several performance measures for risk models, and show how they are related.
\ Application of the measures to risk models for hypothetical populations
and for postmenopausal US\ women illustrate several points. First, model
performance is constrained by the distribution of true risks in the
population. This complicates the comparison of two models if they are
applied to populations with different covariate distributions. \ Second, the
Brier Score and the Integrated Discrimination Improvement (IDI) are more
useful than the concordance statistic for quantifying precision gains
obtained from model expansion. \ Finally, these precision gains are apt to
be small, although they may be large for some individuals. We propose a new
way to identify these individuals, and show how to quantify how much they
gain by measuring the additional covariates. Those with largest gains could
be targeted for cost-efficient covariate assessment. \ 

\bigskip


\bigskip

\bigskip

\begin{center}
\textbf{1. INTRODUCTION}
\end{center}

\noindent \qquad People want to know their risks of future adverse health
outcomes, to help weigh the pros and cons of risk-reducing interventions. \
A person's risk $p$ is determined by his values $z$ of a set of
risk-determining covariates: $p=\xi \left( z\right) .$ Since the full set of
covariates $z$ is seldom known, a \textit{risk model} uses a subset $x$ of
the covariates to assign a risk $r=\gamma \left( x\right) .$ The assigned
risks of a good risk model are \textit{accurate} (i.e., agree with outcome
prevalences within subgroups of the population) and \textit{precise} (i.e.,
able to discriminate those with different true risks). \ \ To evaluate risk
models, we consider three measures of model performance, developed for use
in meteorology, economics and psychology [1-4]. \ The Brier score evaluates
both the accuracy and the precision of assigned risks. In contrast, the
risk-outcome (RO) correlation and the commonly used concordance [5] evaluate
only the precision of a risk model. \bigskip

\noindent \qquad An important issue is how much the precision of a risk
model can be improved by expanding it with additional covariates. The
precision component of the Brier score is particularly useful for resolving
this issue. \ We show how to use it to identify individuals who may benefit
substantially from such expansion, even when the overall gain in the
population is small. \ We also show that precision gains are constrained by
the distribution $\varphi \left( z\right) $ of covariates in the population
to which the original and expanded models are applied. This constraint
complicates comparison of two risk models, especially if they have been
applied to two different populations.

\begin{center}
\bigskip

\textbf{2. POPULATION RISKS AND RISKS MODELS}
\end{center}

\textit{2.1 Population Risks}

\noindent \qquad Consider a population of women aged 50 years who are at
risk of developing breast cancer within a ten year period. Each woman has an
unknown probability $p$ of this outcome, which depends on covariates $z$
that determine her breast cancer hazard rate and her mortality rate. \ The
covariates $z$ may include continuous and/or discrete components, and thus
the risks $p$ may assume continuous or discrete values. In practise,
however, continuous covariates and risks are grouped into finitely many
discrete categories, and here we shall represent all risks as discrete.
\bigskip

\qquad Some of the covariates may be unknown. For example, a woman's risk
may depend on an unknown combination of her genetic inheritance and hormonal
exposures. Women with different covariates $z$ may have the same risk $p$. \
For instance, a low breast cancer risk may pertain to one woman whose
covariates indicate low risks for both breast cancer and competing causes of
death during the period, and to another whose covariates indicate high
breast cancer risk but also high mortality risk. \ \ The population
covariate distribution $\varphi \left( z\right) $ determines the population
risk distribution $f\left( p\right) $\ via the relation%
\begin{equation}
f\left( p\right) =\sum_{z:\xi \left( z\right) =p}\varphi \left( z\right) .
\label{f}
\end{equation}%
The mean $\pi $ of the risk distribution $f$ specifies the prevalence of the
outcome in the population. \ The variance $\sigma ^{2}$ of $f$ specifies the
degree of risk heterogeneity in the population. \ \bigskip

\noindent \qquad Figure 1 shows two risk distributions, both with mean 10\%,
that provide bounds on population risk heterogeneity for all populations
with outcome prevalence $\pi =10\%$. At one extreme, the "constant"
distribution $c$ (left panel)\ assigns all mass to the mean risk $\pi $: 
\begin{eqnarray}
c\left( p\right) &=&1\text{ \ if }p=\pi  \label{cy} \\
&=&0\text{ \ \ else.}  \notag
\end{eqnarray}%
Under this distribution, all in the population have the same risk $\pi $,
and the risk variance $\sigma ^{2}=0.$ Thus the variance of this
distribution gives a lower bound on that of any distribution with mean $\pi $%
. \ At the other extreme, the deterministic distribution $d$ (right panel)
assigns mass $1-\pi $ to $p=0$ and mass $\pi $ to $p=1$: 
\begin{eqnarray}
d\left( p\right) &=&1-\pi \text{ \ \ \ if }p=0  \label{dy} \\
&=&\pi \text{ \ \ \ \ \ \ \ \ \ if }p=1.  \notag
\end{eqnarray}%
Under this distribution, the outcome occurs deterministically, with
probability one in a fraction $\pi $ of the population, and with probability
zero in the remainder. The risk variance is $\sigma ^{2}=\pi \left( 1-\pi
\right) $, which gives an upper bound on the variance of any risk
distribution with mean $\pi .$ When the mean risk is $\pi =10\%$, for
example, the maximum standard deviation of risks is 30\%.\ \bigskip

\qquad Since the variance $\sigma ^{2}=0$ of distribution (\ref{cy}) is a
lower bound on $\sigma ^{2}$, and that of distribution (\ref{dy}) is an
upper bound, we have 
\begin{equation}
0\leq \sigma ^{2}\leq \pi \left( 1-\pi \right) .  \label{bd}
\end{equation}

\textit{2.2 Risk Models}

\noindent \qquad The complete covariates $z$ that determine a woman's breast
cancer risk are not known. Therefore we use measured covariates $x=$\ $%
\kappa \left( z\right) ,$ and outcomes obtained from cohort data to develop
a risk model $\gamma .$ It assigns to all women with covariates $x$ a risk $%
\gamma (x)=r$, $0<r<1.$ Then, given a woman's measured covariates, we
approximate her unknown true risk $p$ by her assigned risk $r$. \ \bigskip

\qquad The distribution of a model's assigned risks is $g\left( r\right)
=\sum \varphi \left( z\right) ,$\ where the summation is taken over $\left\{
z:\gamma \left[ k\left( z\right) \right] =r\right\} ,$ and the joint
distribution of true risk $P$ and assigned risk $R$ is $f_{P,R}\left(
p,r\right) =\sum \varphi \left( z\right) ,$ with summation taken over $%
\left\{ z:\xi \left( z\right) =p\text{ }\&\text{ }\gamma \left[ k\left(
z\right) \right] =r\right\} $. \ The distribution of true risks among
individuals assigned risk $r\ $\ is%
\begin{equation}
f_{P|R}\left( p|r\right) =\frac{f_{P,R}\left( p,r\right) }{g\left( r\right) }%
.  \label{PgivenR}
\end{equation}%
The mean $\pi \left( r\right) $ of this distribution is the outcome
prevalence in the subgroup of individuals with assigned risk $r$ [6]. \
\bigskip

\qquad The variance of the outcome prevalences $\pi \left( r\right) $ across
risk groups equals their covariance with individual outcomes $y$, as shown
in Appendix equation (\ref{cov}). This covariance is bounded by the
covariance between individual outcomes and true risks: 
\begin{equation}
cov_{Y,R}\left[ y,\pi \left( r\right) \right] =var_{R}\left[ \pi \left(
r\right) \right] \leq cov_{Y,P}\left( y,p\right) =\sigma ^{2}.  \label{ineq}
\end{equation}%
Combining inequalities (\ref{bd}) and (\ref{ineq}) gives an ordering for the
variances of outcome prevalences in assigned and true risk groups:%
\begin{equation}
0\leq var_{R}\left[ \pi \left( r\right) \right] \leq \sigma ^{2}\leq \pi
\left( 1-\pi \right) .  \label{vineq}
\end{equation}%
We shall see that the variance of outcome prevalences across a model's
assigned risk groups determines its precision. The inequalities (\ref{vineq}%
) indicate that this variance is bounded by the heterogeneity of true risks
in the population (which the investigator cannot control).\bigskip

\noindent \qquad To illustrate this point with a simple hypothetical
example, consider the risk distributions in panels (a) and (b) of Figure 2
for two populations with appreciably different distributions of
risk-determining covariates $z=z_{0},z_{1},z_{2},z_{3}$. \ Specifically, the
distribution of $z$ depends on a parameter $\alpha $ in the unit interval,
and Population A corresponds to $\alpha =0.2$ while Population B corresponds
to $\alpha =0.8$. \ Both risk distributions have mean $\pi =10\%$, and both
sets of risks range from $2\%$ \ to $74\%$. \ However the variance of risks
is much larger for Population B ($\sigma ^{2}=2.83\%$) than for Population A
($\sigma ^{2}=0.30\%$). \ Now consider a model that assigns risks using only
the first two of the covariates $z_{0},z_{1},z_{2},z_{3}$. Panels (c) and
(d) show the distributions $g\left( r\right) $ of outcome prevalences $\pi
\left( r\right) $ within the five risk groups of this model. \ The variance
of outcome prevalences in Population A is $var_{R}\left[ \pi \left( r\right) %
\right] =0.24\%,$ which by (\ref{vineq}) is bounded by the variance $\sigma
^{2}=0.30\%$ of true risks. \ In contrast, the variance $var_{R}\left[ \pi
\left( r\right) \right] $ in Population B is $2.57\%$, with the more relaxed
upper bound of 2.83\%. Note that, in both populations, omitting covariates
results in rearranging the nine risk groups of panels (a) and (b) into the
five groups of panels (c) and (d), with loss of precision. In particular,
the group in panel (c)\ with outcome prevalence of $33\%$ consists of some
individuals whose true risk is $26\%,$ others whose true risk is $42\%$ and
still others whose true risk is as high as $74\%$. \ (The covariate
distribution $\varphi \left( z\right) ,$ the relation $p=\xi \left( z\right) 
$ and the outcome prevelances $\pi \left( r\right) $ for this example are
given in the Appendix.)

\begin{center}
\bigskip

\textbf{3. PERFORMANCE MEASURES FOR RISK MODELS}
\end{center}

\noindent \qquad What attributes do we want in a risk model? First, we want
the model to assign each individual an accurate risk, i.e., one that agrees
well with his or her true risk. How do we measure accuracy when we don't
know this true risk? Since a person's true risk is the probability of
developing the outcome, his or her assigned risk $r$ should agree with the
outcome prevalence $\pi \left( r\right) $ (i.e., the mean true risk)\ among
all those assigned risk $r$. A model's \textit{calibration} describes this
agreement. \ \bigskip

However a model, even though well-calibrated, should not assign a single
risk to a group of individuals whose true risks vary substantially. Such a
group might consist of two subgroups, one containing individuals at high
risk and the other individuals at low risk. Overall, their outcome
prevalence might agree with their assigned risk, but important
risk-determining covariates would not be reflected in the model. Thus the
second desirable attribute of a model is its \textit{precision} (also called 
\textit{resolution} or \textit{discrimination}), which reflects its ability
to sort the population into subgroups with different true risks. \ \bigskip

\qquad In considering a model's calibration and precision, it is useful to
compare it to a \textit{perfect} model (Model P) that assigns each
individual his or her true risk: $r=p$. \ The risks of Model P are those
people would receive if we could measure their complete covariates $z$, and
could correctly specify the relation $p=\xi \left( z\right) $ between
covariates and risk. \ The distribution of assigned risks for Model P is
just $g\left( r\right) =f\left( r\right) .$

\bigskip

\textit{3.1 Model Calibration }\ 

\noindent \qquad The calibration of a risk model describes how well its
assigned risks agree with outcome prevalences in subgroups of the
population. \ The \textit{calibration bias} in a subgroup of individuals
assigned a given risk $r$ is the difference $r-\pi \left( r\right) $ between
assigned risk and outcome prevalence in the subgroup. \ A risk model is said
to be \textit{well-calibrated} if \ its calibration bias is zero for all
assigned risks $r.$ Calibration biases are displayed in an \textit{%
attributes diagram} [7], which is a plot of the points $\left( r,\pi \left(
r\right) \right) $ for given assigned risks $r.$ For well-calibrated models,
such as the perfect Model P, these points lie on the 45-degree line. \bigskip

\qquad To illustrate calibration, we return to the example of Figure 2. \ A
risk model based on the two covariates $z_{0},z_{1}$ can assign each of its
five subgroups any risk between 0 and 1. However it is well-calibrated to
the population only if it assigns them risks equal to their outcome
prevalences. Thus a model that is well calibrated to Population A assigns
the five risks shown on the abscissa of panel (c) of Figure 2. These are
determined by Appendix formula (\ref{R2}) with the parameter $\alpha =0.2$.
\ If these same risks were assigned to individuals in Population B, the
model would show the calibration biases displayed in the attribute diagram
in the upper panel of Figure 3. These biases occur because Populations A and
B have different distributions of the two unmeasured risk-determining
covariates whose distributions are determined by $\alpha $, and thus they
have different outcome prevalences within assigned risk groups. Note that
the largest biases occur among those at highest risk. \ \bigskip

\noindent \qquad A useful summary of a model's accuracy is its overall
calibration bias, whose square is the average of the squared biases $r-\pi
\left( r\right) $ (vertical distances in Figure 3), weighted by the
proportions $g\left( r\right) $ of individuals in the assigned risk groups: 
\begin{equation}
Bias_{g}^{2}=E_{R}\left\{ \left[ r-\pi \left( r\right) \right] ^{2}\right\}
=\sum\nolimits_{0\leq r\leq 1}g\left( r\right) \left[ r-\pi \left( r\right) %
\right] ^{2}.  \label{Bgbias}
\end{equation}%
The overall calibration bias for the data in the upper panel of Figure 3 is $%
Bias_{g}=8.9\%.$

\bigskip

\textit{3.2 Model Precision}

\qquad We consider three closely related precision measures: the Brier
precision loss, the risk-outcome correlation coefficient, and the
concordance. \ These measures depend on the variances of outcome prevalences 
$\pi \left( r\right) $ in subgroups assigned the same risk $r$. \ They do
not depend on the variance of the actual assigned risks (which may vary
considerably). \ Thus using poorly calibrated assigned risks to measure a
model's precision can be misleading. \bigskip

\qquad A model's \textit{Brier score} [1], also called its mean probability
score [3], is the mean squared error between individual outcomes and risk: 
\begin{equation*}
BS_{g}=E_{Y,R}\left[ \left( y-r\right) ^{2}\right] .
\end{equation*}%
The Brier score measures both the accuracy and the precision of a model. To
see this, we decompose it as \ 
\begin{equation}
BS_{g}=Bias_{g}^{2}+PL_{g}.  \label{BS6}
\end{equation}%
Here $Bias_{g}$ is the model's calibration bias defined by (\ref{Bgbias}),
and 
\begin{equation}
PL_{g}=\pi \left( 1-\pi \right) -var_{R}\left[ \pi \left( r\right) \right]
\label{PLg}
\end{equation}%
is its \textit{precision loss.} $PL_{g}$ is the difference between the
maximum risk variance $\pi \left( 1-\pi \right) $ for a deterministic
outcome and the variance of outcome prevalences $\pi \left( r\right) $
across population subgroups with specific assigned risks $r$. \ The larger
the latter variance, the smaller the precision loss and the more precise the
model.\ The Appendix\ contains a proof of the decomposition (\ref{BS6}), due
to Murphy [2]. \ \bigskip

\noindent \qquad A model's Brier precision loss is closely related to its 
\textit{risk-outcome (RO)\ correlation coefficient}%
\begin{equation}
\rho _{g}=\frac{cov_{Y,R}\left[ y,\pi \left( r\right) \right] }{\sqrt{%
var_{Y}\left( y\right) var_{R}\left[ \pi \left( r\right) \right] }}=\sqrt{%
\frac{var_{R}\left[ \pi \left( r\right) \right] }{\pi \left( 1-\pi \right) }}%
.  \label{rhosq}
\end{equation}%
The second equality in (\ref{rhosq}) follows from (\ref{ineq}).\ \ Combining
(\ref{rhosq}) and (\ref{PLg}) we see that the Brier precision loss is%
\begin{equation}
PL_{g}=\left( 1-\rho _{g}^{2}\right) \pi \left( 1-\pi \right) .  \label{PL9}
\end{equation}%
\ 

\qquad Inequality (\ref{vineq}) shows that a model's precision loss and RO\
correlation are both bounded by their counterparts for the unknown perfect
model. \ In particular, 
\begin{equation*}
0\leq \rho _{g}\leq \rho _{f}\leq 1,
\end{equation*}%
where $\rho _{f}$ is the RO correlation coefficient for the perfect model,
and 
\begin{equation*}
0\leq \pi \left( 1-\pi \right) -\sigma ^{2}\leq PL_{g}\leq \pi \left( 1-\pi
\right) ,
\end{equation*}%
where $\sigma ^{2}$ is the population variance of the true risks. The values
of $\rho _{g}$ and $PL_{g}$ approach their optimal levels as the risk model
captures increasingly many of the risk-determining covariates. \bigskip

\qquad The second and third columns of Table 1 show Brier precision loss and
RO correlation coefficient for a model applied to populations A and B of
Figure 2, using the two covariates $z_{0},z_{1}$ (Model 1). Comparison of
these values with those of the perfect Model P shows that the poor precision
of Model 1 is caused more by homogeneity of true risks than loss of
covariate information. For example, the RO correlation coefficients for
Model 1 (16.3\% for Population A and 53.4\% for Population B) are not far
from their optimal values 18.1\% and 56.1\% for these two populations, given
by the perfect model. \bigskip

\noindent \qquad The most widely used precision measure for health risk
models is the \textit{concordance,} also called the area under the receiver
operating characteristic curve, or area-under-the-curve (AUC)\textit{.} To
describe it, we introduce the conditional distributions of assigned risk
among those who do and do not develop the outcome:%
\begin{equation}
h_{y}\left( r\right) =\Pr \left( R=r|Y=y\right) =\frac{g\left( r\right) %
\left[ \pi \left( r\right) \right] ^{y}\left[ 1-\pi \left( r\right) \right]
^{1-y}}{\pi ^{y}\left( 1-\pi \right) ^{1-y}},\text{ \ \ \ \ }y=0,1.
\label{gy}
\end{equation}%
The concordance is the probability that a risk chosen from the distribution $%
h_{1}$ exceeds one chosen from $h_{0}$: 
\begin{equation}
\zeta _{g}=\sum\nolimits_{0\leq r\leq 1}H_{1}\left( r\right) h_{0}\left(
r\right)  \label{conc}
\end{equation}%
[5,8]. \ Here the cumulative distribution $H_{1}\left( r\right) =\frac{1}{2}%
h_{1}\left( r\right) +\sum_{u>r}h_{1}\left( u\right) $ when $h$ is discrete,
and $H_{1}\left( r\right) =\int_{r}^{1}h_{1}\left( u\right) du$ when $h$ is
continuous. \ The concordance is invariant under any rank-preserving
transformation of the assigned risks. \ Thus any set of assigned risks whose
ranks agree with those of the outcome prevalences $\pi \left( r\right) $ has
the same concordance as that of a well-calibrated risk model. \ The
concordance achieves its minimum of $50\%$ when the risk distribution is the
constant one $c\left( p\right) $ of (\ref{cy})$,$ and its maximum of $100\%$
when the model is perfect and the true risk distribution $f\left( p\right)
=d\left( p\right) $ of (\ref{dy}). \bigskip

\qquad As seen in Table 1, the concordance of Model 1 is 59.4\% for
Population A and 80.6\% for Population B. For comparison, concordances of
the perfect model for these two populations are 60.3\% and 83.3\%,
respectively. \bigskip

\noindent \qquad Since the concordance is a function of the conditional
distributions $h_{1}\left( r\right) $ and $h_{0}\left( r\right) $, it is
most easily interpreted as a retrospective assessment of how well a model
discriminates the risks of those with and without the outcome. \ Indeed, the
model's Brier precision loss and RO\ correlation coefficient also can be
interpreted this way. To see this, note from (\ref{gy}) and (\ref{rhosq})
that the squared RO correlation coefficient can be written as the mean of
the outcome prevalences $\pi \left( r\right) ,$ averaged over the assigned
risk subgroups of those who develop the outcome, minus the corresponding
mean in those who do not:%
\begin{eqnarray}
\rho _{g}^{2} &=&E_{R|Y}\left[ \pi \left( r\right) |y=1\right] -E_{R|Y}\left[
\pi \left( r\right) |y=0\right]  \notag \\
&\equiv &\sum\nolimits_{r}\pi \left( r\right) \left[ h_{1}\left( r\right)
-h_{0}\left( r\right) \right] .  \label{ID}
\end{eqnarray}%
This representation has motivated the name \textit{Integrated Discrimination 
}for $\rho _{g}^{2}$ [9]. \ The Brier precision loss shares this
retrospective interpretation, since it is proportional to $1-\rho _{g}^{2}$
by (\ref{PL9}). However the interpretation is not particularly useful for
health risk assessment, whose goal is to classify individuals prospectively
into risk strata requiring different preventive strategies [10-12]. \ To
their advantage, the Brier precision loss and RO correlation also can be
interpreted prospectively as measures of the extent to which individuals
with different true risks are grouped together and assigned a common risk.
This interpretation has heuristic value for individual risk assessment [13].

\bigskip

\textit{3.3 Comparing Different Risk Models}

\noindent \qquad Let $x_{1}$ and $x_{2}$ represent two sets of risk-related
covariates measured on the same population, and suppose we have developed
risk models based on each of them, with assigned risks $\gamma _{1}\left(
x_{1}\right) =r_{1}$ and $\gamma _{2}\left( x_{2}\right) =r_{2}$ (Models 1
and 2 respectively). \ Often $x_{2}$ consists of $x_{1}$ plus additional
covariates. For example, the covariates $x_{1}$ might consist of known
breast cancer risk factors [14] and the covariates $x_{2}$ might augment $%
x_{1}$ with breast density measurements [15] or the genotypes of certain
susceptibility variants [16]. \ An important issue is how to compare the
performances of the two models, which may differ in both calibration
accuracy and precision. \bigskip

\qquad The Brier score is particularly useful for such comparison. To see
this, we use (\ref{BS6}-\ref{PLg}) to write the difference in Brier scores
for two models as%
\begin{eqnarray}
BS_{g_{1}}-BS_{g_{2}} &=&\left( Bias_{1}^{2}-Bias_{2}^{2}\right) +\left(
PL_{g_{1}}-PL_{g_{2}}\right)  \notag \\
&=&\left( Bias_{1}^{2}-Bias_{2}^{2}\right) +\left\{ var_{R_{2}}\left[ \pi
\left( r_{2}\right) \right] -var_{R_{1}}\left[ \pi \left( r_{1}\right) %
\right] \right\} .  \label{ttt}
\end{eqnarray}%
Here the subscripts 1 and 2 refer to the two risk models. \ The first
summand of (\ref{ttt}) is the gain or loss of squared calibration bias in
Model 2 compared with Model 1. The second summand, called the \textit{Brier
precision difference,} is the variance of outcome prevalences across
subgroups assigned common risks by Model 2, minus the corresponding variance
for Model 1. From (\ref{PL9}) we see that it is proportional to the
difference in the two models' squared RO correlation coefficients:%
\begin{equation}
PL_{1}-PL_{2}=var_{R_{2}}\left[ \pi \left( r_{2}\right) \right] -var_{R_{1}}%
\left[ \pi \left( r_{1}\right) \right] =\pi \left( 1-\pi \right) \left( \rho
_{g_{2}}^{2}-\rho _{g_{1}}^{2}\right) .  \label{PL8}
\end{equation}%
The difference $\rho _{g_{2}}^{2}-\rho _{g_{1}}^{2}$ has been called the 
\textit{Integrated Discrimination Improvement }(IDI) by Pencina et al
(2008), who propose it to measure the precision gained by adding covariates
to a model. \ Both the IDI and the Brier precision difference compare the
models' abilities to sort individuals into groups having different true
risks.\bigskip

\qquad \textit{Comparing calibration}. We illustrate the calibration
component of the Brier score difference (\ref{PL8}) using the two
populations in Figure 2 whose true risks are determined by the four
covariates $z_{0},z_{1},z_{2},z_{3}$. Suppose we expand a model based on $%
z_{0},z_{1}$ (Model 1) by including the additional covariate $z_{2}$. Panels
(e) and (f) of Figure 2 show the distributions of outcome prevalences within
the seven assigned risk groups of the expanded model (Model 2). These are
given by Appendix equation (\ref{R3}). \ Consider the calibration biases
that would result if Models 1 and 2 were each well-calibrated to Population
A but applied to Population B, as shown in Figure 3. \ The overall
calibration bias for Model 2 is $5.5\%$ (panel (b)), an improvement in
accuracy compared to the value $8.9\%$ for Model 1 (panel (a)). In general,
however, expanding a risk model with additional covariates need not reduce
its calibration bias. In Population B, for instance, Model 2 might assign
biased risks to each of its two risk groups having outcome prevalences $%
38.8\%$ and $67.6\%$ (panel (f)\ in Figure 2), whereas Model 1 might assign
a well-calibrated risk of $61.8\%$ to the combined group in panel (d).
\bigskip

A more informative comparison of the calibration biases of two risk models
for a population would focus on individuals assigned a common risk by each
of the two models. To make such a comparison, we embed both models in a
common well-calibrated model, which we call their \textit{cross-classified
model}. \ This model partitions the population into subgroups of individuals
who receive a common assigned risk from each of the two models. To describe
it, consider in Figure 4 how Population B is partitioned into five risk
groups by a model based on two covariates (Model 1) and into seven risk
groups by an expanded model that uses an additional covariate (Model 2).
Each risk group (denoted by a pie wedge) is labeled with its outcome
prevalence. \ Note that some individuals assigned different risks by Model 1
receive the same risk by Model 2. This shows that adding covariates to a
model need not yield a more refined partition of the population. To obtain
the cross-classified model (Model C), we cross-classify individuals
according to the risks they receive by the two models, and then assign each
of the resulting subgroups a common risk equal to the outcome prevalence in
that group. Thus the cross-classified model assigns to those with risk $%
r_{1} $ from Model 1 and risk $r_{2}$ from Model 2 the risk $\pi \left(
r_{1},r_{2}\right) $ which is their outcome prevalence (i.e., their mean
true risk). \ Figure 4 also shows the partition of Population B induced by
the cross-classified model for Models 1 and 2. \bigskip

\qquad We can now compare the models' biases in each of the nine risk groups
of the cross-classified model. Figure 5 shows the outcome prevalences in
these nine risk groups, and the risks assigned by the two models when each
is well-calibrated to Population A but applied to Population B. \ Both
models exhibit substantial downward bias in the group with the highest
outcome prevalence, with at most small bias in the five groups at lowest
risk. Moreover compared to Model 1, the expanded Model 2 is less biased in
groups 7 and 9, but more biased in groups 6 and 8. \bigskip

\qquad \textit{Comparing precision}. \ While expanding a model with
additional covariates may actually decrease its overall accuracy, in
general, such expansion will increase its precision. \ The issue is whether
the precision increase is large enough to warrant the cost of measuring and
modeling the additional covariates. \ To illustrate this point, consider the
precision measures shown in Table 1 for models using two and three
covariates of the populations of Figure 2 (Models 1 and 2 respectively).
Note that, relative to Model 1, Model 2 yields only small gains in precision
for individuals in Population A: the Brier precision difference is only $%
0.03\%$, the RO correlation coefficient increases by only $0.9\%$, and the
concordance increases by only $0.5\%$. The gains are slightly larger for the
more heterogeneous Population B: \ $0.13\%$ in Brier precision difference, $%
1.3\%$ in RO correlation, and $1.5\%$ in concordance. \ \ Nevertheless, the
overall precision gain from expanding Model 1 with additional covariates is
small by any measure, for each population. \ \bigskip

\qquad This example shows that the overall precision gain can be negligible,
even though the expanded model may be considerably more precise in certain
subgroups of the population [12]. \ This anomaly occurs because individuals
with small risks, who typically comprise most of the population, tend to
benefit little from increased precision, while the few with large risks have
much to gain. Since the precision measures are averages over the entire
population, the small gains for the majority tend to dominate the larger
gains for a few. This problem can be addressed by creating subgroup-specific
measures of relative precision, where each subgroup consists of individuals
assigned a common risk by a given model (say, Model 1).\bigskip

\qquad We use the Brier precision difference to describe how to find the
Model 1 risk groups that gain the most. The idea is to embed the two models
in their common well-calibrated cross-classified model, and then use the
latter to evaluate the increased precision for each risk group of Model 1.
Specifically, we write the overall Brier precision difference (\ref{PL8})
between Models 1 and 2 in terms of differences from their common
cross-classified model: 
\begin{equation*}
PL_{1}-PL_{2}=\left( PL_{1}-PL_{E}\right) -\left( PL_{2}-PL_{E}\right) .
\end{equation*}%
Equation (\ref{PL8}) shows that the precision difference $PL_{1}-PL_{E}$ is
the variance of outcome prevalences across risk groups determined by the
cross-classified model, minus the corresponding variance for groups
determined by Model 1:%
\begin{equation*}
PL_{1}-PL_{E}=var_{R_{1},R_{2}}\left[ \pi \left( r_{1},r_{2}\right) \right]
-var_{R_{1}}\left[ \pi \left( r_{1}\right) \right] .
\end{equation*}%
We show in the Appendix that this difference is the average of
subgroup-specific variances: 
\begin{eqnarray}
PL_{1}-PL_{E} &=&E_{R_{1}}\left\{ var_{R_{1},R_{2}|R_{1}}\left[ \pi \left(
r_{1},r_{2}\right) |r_{1}\right] \right\}  \notag \\
&=&\sum_{0\leq r_{1}\leq 1}g_{1}\left( r_{1}\right) var_{R_{1},R_{2}|R_{1}} 
\left[ \pi \left( r_{1},r_{2}\right) |r_{1}\right] .  \label{s}
\end{eqnarray}%
The variance of outcome prevalences within the subgroup assigned a given
risk $r_{1},$ denoted $var_{R_{1},R_{2}|R_{1}}\left[ \pi \left(
r_{1},r_{2}\right) |r_{1}\right] $ in (\ref{s}), gives a measure of
precision gain in this risk group. \ These variances (or their corresponding
standard deviations) can be used to determine those risk groups for whom
measuring additional covariates has the greatest benefit.\bigskip

\qquad To illustrate this assessment, suppose we wish to determine the Model
1 risk groups in Population B who gain most from the additional covariate
used by Model 2. Table 2 shows the ranges and standard deviations of outcome
prevalences in the nine subgroups determined by the cross-classified model,
within each of the five risk groups of population B assigned a common risk
by Model 1. The precision gains, as measured by these standard deviations,
range from zero (in those assigned a risk of 10\% by Model 1) to 11.5\% (in
those assigned the highest risk of 61.8\%). Thus the individuals at highest
risk gain substantially more precision from the additional covariate than do
other individuals. If this covariate were difficult to measure, it might be
deemed cost-efficient to do so only for those individuals to whom Model 1
assigns the highest risks.

\begin{center}
\bigskip

\textbf{4. APPLICATION TO DATA}
\end{center}

We illustrate the performance measures by application to two models for risk
of estrogen-receptor-positive (ER+) breast cancer among postmenopausal
women. These models were developed by Rosner et al [17] using data from the
prospective Nurses Health Study (NHS). \ The investigators identified 1559
ER+ breast cancer cases in 476,581 person-years of followup among
postmenopausal women with a natural menopause, giving a crude annual
incidence rate of 1559/476,581 = 0.0021 cases per woman per year. \ To
illustrate the methods described here, we consider the ten-year breast
cancer risks induced by this incidence rate in a hypothetical cohort of
postmenopausal women aged 50 years with the same covariate distribution as
the NHS\ women. \ \ The annual death rate for US white women aged 50-59
years is 0.0053 deaths per woman per year [18]. \ This rate, combined with
the breast cancer incidence rate of 0.0021, yields a mean risk $\pi =2.02\%$ 
$\ $of developing ER+ breast cancer within ten years. \ Because this risk is
low, the maximum risk heterogeneity in this population is also low: the
standard deviation of risks is bounded above by $\sqrt{.0202\left(
1-.0202\right) }=14.1\%.$\bigskip

\ \ \ \ \qquad Rosner et al [17] developed a risk model based on each
woman's current age, age at menarche, age at natural menopause, parity, ages
at all births, history of benign breast disease, history of breast cancer in
a mother or sister, and years of use of estrogen replacement therapy (Model
1). \ The investigators also expanded this model by including one additional
covariate representing estimated serum levels of endogenous estradiol (Model
2). \ Table 4 of Rosner et al [17] presents case counts, person-years of
followup, and incidence rates within subgroups of the population
cross-classified by joint deciles of risk as assigned by Models 1 and 2.
Here we used these incidence rates and all-cause mortality rates to compute
the ten-year breast cancer risks shown in Table 3. \bigskip

\qquad Models 1 and 2 were calibrated to the NHS population, and we do not
investigate their potential biases if applied to a similar but independent
cohort of US postmenopausal women. \ Instead we assume that the
probabilities shown in the joint risk groups of Table 3 represent the mean
true risks, i.e. the ten-year breast cancer prevalences in each joint risk
group. \ With this assumption, the 40 nonempty cells in the interior of
Table 3 represent the joint risk groups of the cross-classified model for
Models 1 and 2. \ From the row and column margins of Table 3 we calculate a
standard deviation for Model 1 risks of $\left\{ \sum_{r_{1}=1}^{10}0.1\left[
\pi \left( r_{1}\right) -0.0202\right] \right\} ^{1/2}=1.04\%$, while that
for Model 2 risks is $1.22\%$. These values correspond to RO correlation
coefficients of 3.1\% and 4.6\%, respectively, and to a small overall Brier
precision gain of $.004\%$. \ \bigskip\ 

\qquad Which risk groups of Model 1 benefit most from the addition of
endogenous estradiol levels to the model? \ To answer this question, we
refer to the last two columns of Table 3, which show the range and standard
deviation of the cross-classified outcome prevalences within each of the ten
Model 1 risk groups. \ The smallest range is $0.7\%-1.0\%$ in the first
decile of Model 1, and the largest is $0.6\%-6.5\%$ in the seventh decile.
The standard deviations vary tenfold from $0.12\%$ in group 1 to $1.24\%$ in
group 9. (To calculate these standard deviations, we assumed that within
each decile of Model 1 risk, women were distributed in the same proportions
as the person-years of followup given by Rosner et al [17].) In this
example, there are modest benefits in prediction precision that these women
might receive from adding a serum estradiol measurement. Moreover, the
absolute benefits do not vary appreciably across risk groups determined by
Model 1. \ These results are due in part to the limited range of risks in
this population, and may also reflect measurement error in the estimated
estradiol levels.

\begin{center}
\textbf{5. DISCUSSION}
\end{center}

\noindent \qquad We have described the Brier score, the risk-outcome (RO)
correlation, and the concordance as performance measures for risk models
that use personal covariates to assign personal risks of a future adverse
outcome. The Brier score is the sum of terms representing calibration bias
plus precision loss, while the RO correlation coefficient and concordance
measure only precision loss. Application of the measures to risk models for
two hypothetical populations and for postmenopausal women at risk of breast
cancer illustrates several points.\bigskip

\qquad First, though an individual wanting to know his or her own personal
risk for an adverse outcome is not concerned with how a model performs for
others in his population, paradoxically, model performance is sensitive to
the distribution of covariates in the entire population. \ No risk model can
be more precise than the ideal perfect model that assigns the true risks to
all in the population. In particular, a homogeneous distribution of true
risks induces poor precision values for all risk models for that population,
as noted by Cook [10]. \ Because of this strong dependence of precision
measures on the population risk distribution, comparing summary measures of
model performance in different populations requires caution. For example,
Table 1 shows that if we applied Model 1 to Population B and Model 2 to
Population A, we would conclude that Model 1 is superior to Model 2 by any
precision measure, despite the consistently better precision of Model 2 when
both models are applied and calibrated to the same population. \bigskip

\qquad Evaluating precision in a single population also warrants caution if
the models are not well calibrated. This is because precision reflects how
outcome prevalences vary across the risk groups, and not how the assigned
risks vary. Thus precision measures based on the variances of poorly
calibrated assigned risks can be misleading. \bigskip

\noindent \qquad The calibration accuracy and precision of two risk models
for a given population can be compared using the difference in their values
for an appropriate performance measure. \ The Brier score difference is
particularly useful because it is the sum of a component that compares the
accuracy of assigned risks, and a component that compares their precision.
Each of these two components has a meaningful prospective interpretation for
individuals seeking to know their own risks for a future adverse outcome.
\bigskip

\noindent \qquad Nevertheless, all summary measures of model performance
have two limitations. \ First, because they depend on the population's
distribution of measured covariates, they cannot be compared reliably across
populations with different covariate distributions. \ Second, the precision
gained by adding covariates to a model may not be uniform across all
population subgroups. It may be large in certain subgroups, while the
overall gain may be small. \ Both these limitations can be addressed by
focusing on subgroup-specific performance measures. \ We have proposed a new
method for evaluating subgroup-specific gains or losses in both calibration
and precision by using one model compared to another. \ In particular, we
propose evaluating the precision gain within each risk group defined by one
of the models, by quantifying the spread of outcome prevalences across those
cross-classified according to joint risks from both models. \ This idea is
similar in flavor to that of Cook [19], who proposed evaluating the
percentages of people who move to a different risk quantile after additional
covariate assessment. \ Knowing which subgroups gain the most accuracy and
precision in assigned risk can be useful for preventive efforts, because
these are the people with most to gain by enlarging the risk model to
include the covariates that distinguish their risks. \ This opportunity to
evaluate subgroup-specific performance improvement suggests the following
strategy: evaluate gains within assigned risk groups to determine those
groups most likely to benefit from gathering additional covariate
information; then target them for such covariate assessment. \bigskip

\noindent \qquad Further work is needed to develop and evaluate all personal
risk models [20]. For example, we need models and performance measures that
accommodate multiple correlated adverse outcomes. \ At present, outcomes are
studied in isolation, although preventive interventions typically affect
risks for several outcomes [21]. In addition, most models assume that all
individuals have the same age-specific mortality rates for competing causes
of death, despite large variation in these rates among the elderly [22]. \
These issues should be addressed in future risk models and performance
measures. Finally, we need to understand how risk information is perceived
by patients and their physicians. Small improvements in the precision of
assigned risks are of little significance for personal preventive decisions,
so we need to target those with the largest potential gains. \bigskip

\begin{center}
\textbf{APPENDIX}\bigskip
\end{center}

\noindent \qquad \textbf{Risk distributions of Figure 2}. The risks shown in
panels (a) and (b)\ of Figure 2 are determined by four covariates $%
z_{0},z_{1},z_{2},z_{3}$, with $z_{0}$ coded as $-1,0,1,$ and $z_{j}$ coded
as $0,1,$ $j=1,2,3.$ These covariates determine risk according to the rule%
\begin{eqnarray}
p=\xi \left( z\right) &=&0.1+0.01z_{0}2^{z_{1}+z_{2}+z_{3}},\text{ }%
z_{0}=-1,0,  \label{R1} \\
&=&0.1+0.08z_{0}2^{z_{1}+z_{2}+z_{3}},\text{ }z_{0}=1.  \notag
\end{eqnarray}%
The distributions $f\left( p\right) $ in the two panels are determined by
the relation (\ref{f}), where the covariate distribution $\varphi \left(
z\right) =\tau \left( z_{0}\right) \prod\nolimits_{i=1}^{3}$ $\alpha
^{z_{i}}\left( 1-\alpha \right) ^{1-z_{i}}$. \ Here $\tau \left(
z_{0}\right) =0.8,$ $0.1,$ $0.1,$ for $z_{0}=$ $-1,0,1,$ respectively. \ The
value $\alpha =0.2$ gives the distribution for Population A and $\alpha =0.8$
gives the one for Population B. \ The mean risk is $10\%$ and the variance
is $7.2\times 10^{-4}\left( 1+3\alpha \right) ^{3}.$\bigskip

\qquad Panels (c) and (d) of Figure 2 show outcome prevalences within
population subgroups assigned common risks based on two of the four
covariates: $z_{0},z_{1}$ . These prevalences are%
\begin{equation}
\pi \left[ r_{1}\left( z_{0},z_{1}\right) \right] =\left\{ 
\begin{array}{ccc}
0.1-.01\left( 1+\alpha \right) ^{2}\times 2^{z_{1}}, &  & z_{0}=-1 \\ 
0.1, & \text{if } & z_{0}=0 \\ 
0.1+.08\left( 1+\alpha \right) ^{2}\times 2^{z_{1}}, &  & z_{0}=1%
\end{array}%
\right. \text{. \ }  \label{R2}
\end{equation}

\noindent \qquad Panels (e) and (f)\ of Figure 2 give outcome prevelances in
the seven subgroups determined by expanding a risk model based on $%
z_{0},z_{1}$ (Model 1) to include $z_{2}$. \ These prevalences are%
\begin{equation}
\pi \left[ r_{1}\left( z_{0},z_{1},z_{2}\right) \right] =\left\{ 
\begin{array}{ccc}
0.1-.01\left( 1+\alpha \right) \times 2^{z_{1}+z_{2}}, &  & z_{0}=-1 \\ 
0.1, & \text{if } & z_{0}=0 \\ 
0.1+.08\left( 1+\alpha \right) \times 2^{z_{1}+z_{2}}, &  & z_{0}=1%
\end{array}%
\right. .  \label{R3}
\end{equation}

\noindent \qquad \textbf{Proof of relations (\ref{ineq})}. \ Since $E_{R}%
\left[ \pi \left( r\right) \right] =\pi =E_{Y}\left[ y\right] ,$ we have 
\begin{eqnarray}
cov_{Y,R}\left[ y,\pi \left( r\right) \right] &=&E_{Y,R}\left[ y\pi \left(
r\right) \right] -\pi ^{2}  \notag \\
&=&E_{R}\left\{ E_{Y|R}\left[ y\pi \left( r\right) |r\right] \right\} -\pi
^{2}  \notag \\
&=&E_{R}\left\{ \left[ \pi \left( r\right) \right] ^{2}\right\} -\pi
^{2}=var_{R}\left[ \pi \left( r\right) \right] .  \label{cov}
\end{eqnarray}%
Also, 
\begin{equation*}
\sigma ^{2}=E_{R}\left[ \sigma ^{2}\left( r\right) \right] +var_{R}\left[
\pi \left( r\right) \right] \geq var_{R}\left[ \pi \left( r\right) \right] ,
\end{equation*}%
which implies \textbf{(\ref{ineq}).} \bigskip

\noindent \qquad \textbf{Proof of Brier score decomposition (\ref{BS6})}. We
write%
\begin{eqnarray}
BS_{g} &=&E_{Y,R}\left[ \left( y-r\right) ^{2}\right] =E_{Y,R}\left[ \left(
y-\pi \left( r\right) +\pi \left( r\right) -r\right) ^{2}\right]  \notag \\
&=&E_{R}E_{Y|R}\left[ \left( y-\pi \left( r\right) \right) ^{2}\right]
+2E_{R}\left\{ \left[ \pi \left( r\right) -r\right] E_{Y|R}\left[ y-\pi
\left( r\right) \right] \right\}  \notag \\
&&+E_{R}\left[ \left( \pi \left( r\right) -r\right) ^{2}\right] .  \label{BS}
\end{eqnarray}%
Here 
\begin{eqnarray*}
E_{Y|R}\left[ \left( y-\pi \left( r\right) \right) ^{2}\right]
&=&E_{P|R}E_{Y|P}\left[ y-2y\pi \left( r\right) +\pi ^{2}\left( r\right) %
\right] \\
&=&E_{P|R}\left[ p-2p\pi \left( r\right) \right] +\pi ^{2}\left( r\right) \\
&=&\pi \left( r\right) -\pi ^{2}\left( r\right) .
\end{eqnarray*}%
Thus the first summand in (\ref{BS}) is 
\begin{eqnarray}
E_{R}E_{Y|R}\left[ \left( y-\pi \left( r\right) \right) ^{2}\right] &=&E_{R}%
\left[ \pi \left( r\right) \right] -E_{R}\left[ \pi ^{2}\left( r\right) %
\right]  \notag \\
&=&\pi -\pi ^{2}-var\left[ \pi \left( r\right) \right] .  \label{r}
\end{eqnarray}%
The second summand in (\ref{BS}) vanishes, since 
\begin{equation}
E_{Y|R}\left[ y-\pi \left( r\right) \right] =E_{P|R}E_{Y|P}\left[ y-\pi
\left( r\right) \right] =E_{P|R}\left[ p-\pi \left( r\right) \right] =0.
\label{rr}
\end{equation}%
Substituting (\ref{r}) and (\ref{rr}) into (\ref{BS}) gives (\ref{BS6}%
).\bigskip

\noindent \qquad \textbf{Proof of equation (\ref{s}). }\ By the standard
decomposition of a variance into the expectation of a conditional variance
plus the variance of a conditional expectation, we have 
\begin{eqnarray}
var_{R_{1},R_{2}}\left\{ \pi \left( r_{1},r_{2}\right) \right\} &=&E_{R_{1}} 
\left[ var_{R_{1},R_{2}|R_{1}}\left\{ \pi \left( r_{1},r_{2}\right)
|r_{1}\right\} \right] +var_{R_{1}}\left\{ E\left[ \pi \left(
r_{1},r_{2}\right) |r_{1}\right] \right\}  \notag \\
&=&var_{R_{1}}\left\{ \pi \left( r_{1}\right) \right\} +E_{R_{1}}\left[
var_{R_{1},R_{2}|R_{1}}\left\{ \pi \left( r_{1},r_{2}\right) |r_{1}\right\} %
\right] .  \label{tt}
\end{eqnarray}%
This gives (\ref{s}).\textbf{\ }

\bigskip

\textbf{Acknowledgements}: This research was supported by NIH grant
CA094069. \ I am grateful to Joseph B. Keller for useful discussions and
Nicole Ng for help with the calculations.\pagebreak

\begin{center}
\bigskip \textbf{REFERENCES}
\end{center}

\noindent

\begin{enumerate}
\item Brier GW. Verification of forecasts expressed in terms of probability.
Monthly Weather Review 1950; 78:1-3.

\item Murphy AH. A new vector partition of the probability score. Journal of
Applied Meteorology 1973; 12:595-600.

\item Yates JF. External correspondence: Decompositions of the mean
probability score. Organizational Behavior and Human Performance 1982;
30:132-156.

\item Wilks DS. Statistical methods in the Atmospheric Sciences. London,
Academic Press 1995;

\item Hanley JA, Mcneil BJ. The meaning and use of the area under a receiver
operating characteristic (ROC) curve. Radiology 1982; 143:29-36.

\item Gail MH, Pfeiffer RM. On criteria for evaluating models of absolute
risk. Biostatistics 2005; 6: 227-239.

\item Hsu WR, Murphy AH. The attributes diagram A geometrical framework for
assessing the quality of probability forecasts. International Journal of
Forecasting 1986; 2:285-293.

\item Pepe MS. The Statistical Evaluation of Medical Tests for
Classification and Prediction. Oxford University Press 2003.

\item Pencina MJ, D'Agostino RB Sr., D'Agostino RB Jr., Vasan RS. Evaluating
the added predictive ability of a new marker: from area under the ROC curve
to reclassification and beyond. Statistics in Medicine 2008; 27:157-172.

\item Cook NR. Use and misuse of the receiver operating characteristic curve
in risk prediction. Circulation 2007; 115:928-935.

\item Ridker PM, Buring JE, Rifar N, Cook N. Development and validation of
improved algorithms for the assessment of global cardiovascular risk in
women. Journal of the American Medical Association 2007; 297:611-619.

\item Pepe MS, Janes HE. Gauging the performance of SNPs, biomarkers, and
clinical factors for predicting risk of breast cancer. Journal of National
Cancer Institute 2008; 100:978-979.

\item Graf E, Schmoor C, Sauerbrei W, Schumacher M. Assessment and
comparison of prognostic classification schemes for survival data.
Statistics in Medicine 1999; 18:2529-2545.

\item Gail MH, Brinton LA, Byar DP, Corle DK, Green SB, Schairer C,
Mutvihill JJ. Projecting individualized probabilities of developing breast
cancer for white females who are being examined annually. Journal of
National Cancer Institute 1989; 81:879-886.

\item Chen J, Pee D, Ayyagari R, Graubard B, Schairer C, Byrne C, Benichou
J, Gail MH. Projecting absolute invasive breast cancer risk in white women
with a model that includes mammographic density. Journal of National Cancer
Institute 2006; 98:1215-1226.

\item Gail MH. Discriminatory accuracy from single-nucleotide polymorphisms
in models to predict breast cancer risk. Journal of National Cancer
Institute 2008; 100:1037-1041.

\item Rosner B, Colditz GA, Iglehart JD, Hankinson SE. Risk prediction
models with incomplete data with application to prediction of estrogen
receptor-positive breast cancer: prospective data from the Nurses' Health
Study. Breast Cancer Res 2008; 10:R55.

\item Arias E. United States Life Tables 2004. Natl Vital Stat Report 2007;
56:1-39.

\item Cook NR. Statistical evaluation of prognostic versus diagnostic
models: beyond the ROC curve. Clinical Chemistry 2008; 54:17-23.

\item Pepe MS, Feng Z, Anes H, Bossuyt PM, Potter JD. Pivotal evaluation of
the accuracy of a biomarker used for classification or prediction: standards
for study design. Journal of National Cancer Institute 2008; 100:1432-1438.

\item Giordano SH, Hortobagyi GN. Time to remove the subspecialty blinders:
breast cancer does not exist in isolation. Journal of National Cancer
Institute 2008; 100: 230-231.

\item Walter LC, Covinsky KE. Cancer screening in elderly patients: a
framework for individualized decision making. Journal of the American
Medical Association 2001; 285:2750-2756.
\end{enumerate}



\section*{Supplementary material}

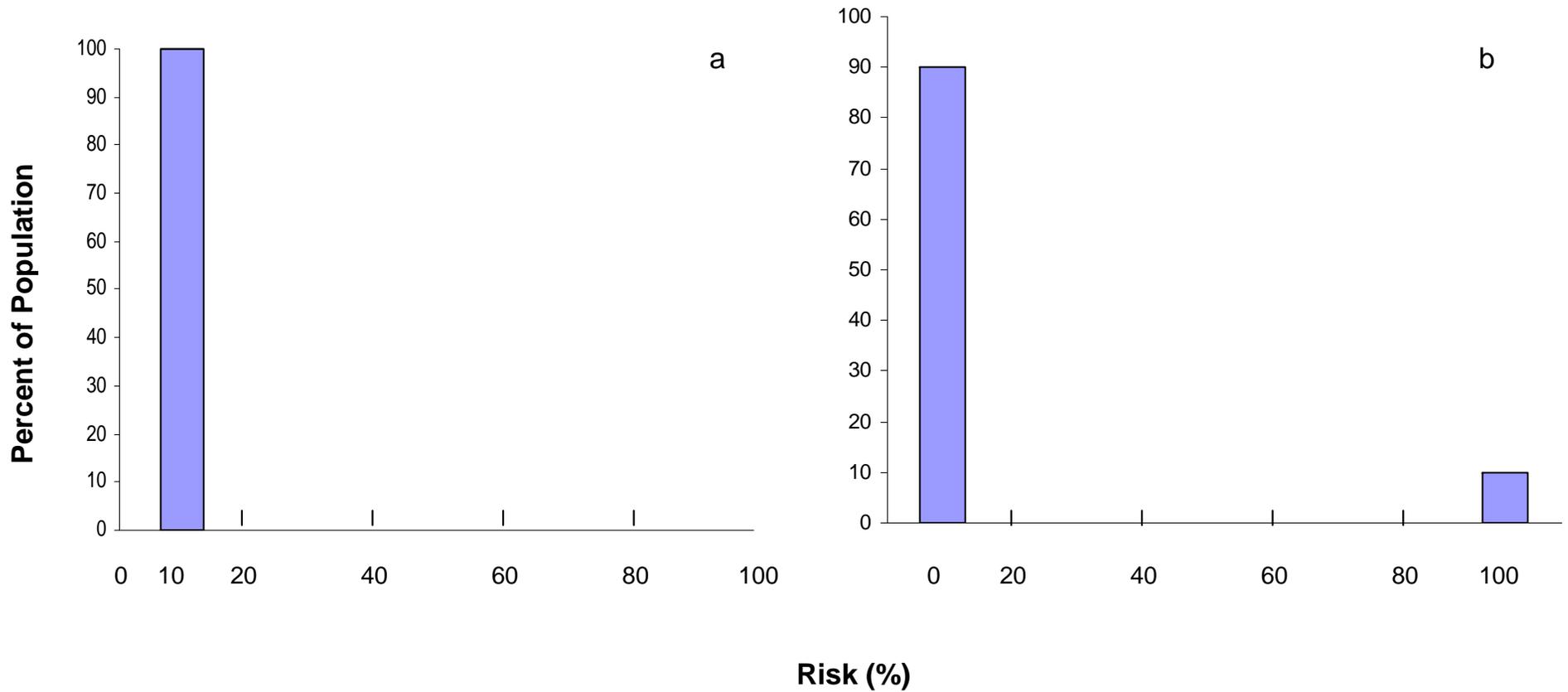

**Plate 1.** Two extreme distributions for a population with outcome prevalence 10%. a) all individuals have common risk of =10%, so that the risk variance is ²=0. b) ninety percent of individuals have risk 0 and the remaining ten percent have risk one. The variance of risks is ²=.10×(1-.10)=.09.

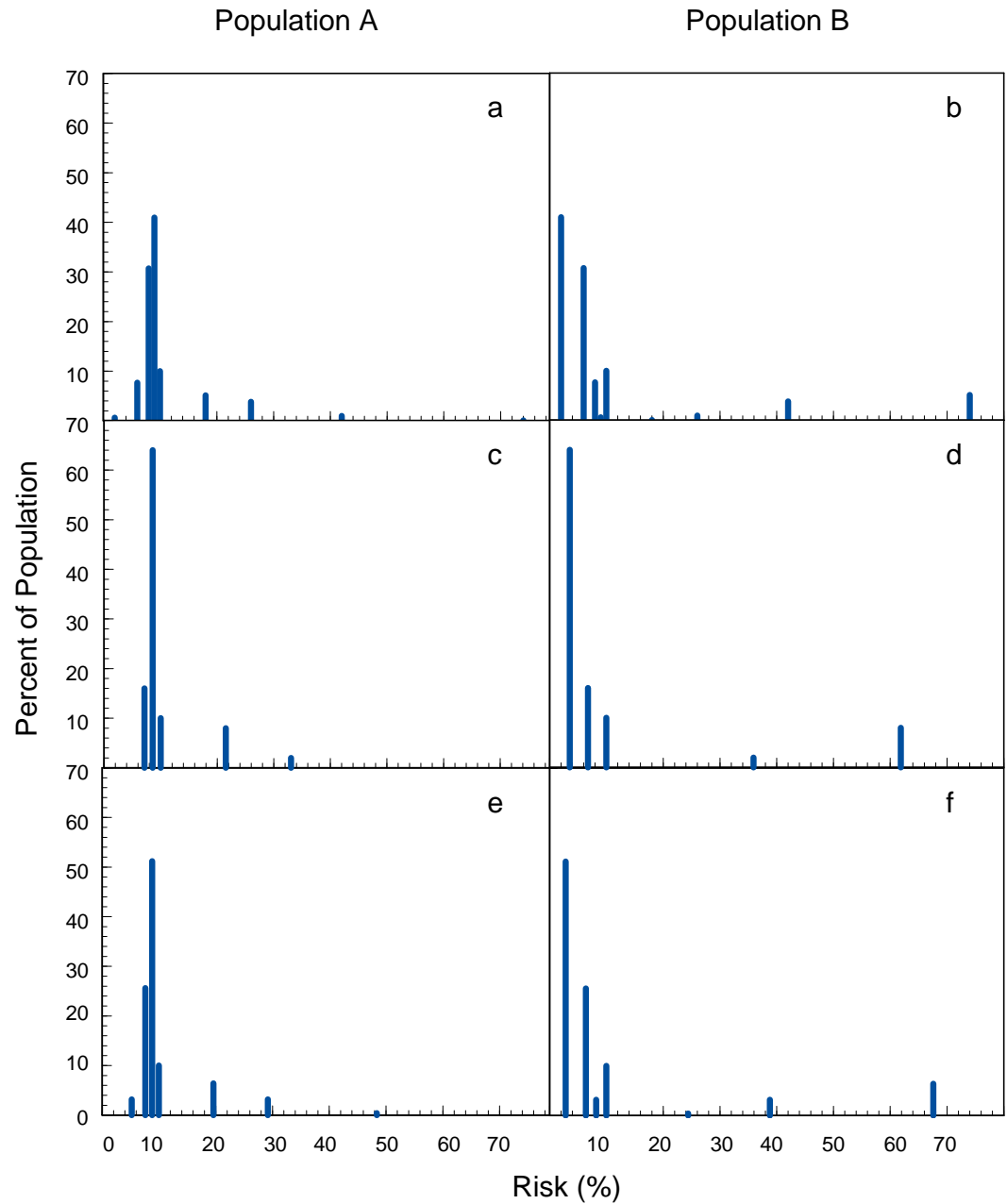

**Plate 2**. Risk distributions for two hypothetical populations, A and B. Risks depend on four covariates. Panels (a) and (b) show the distributions of true risk, with mean =10%. The standard deviation of true risks is 5.4% for population A and 16.7% for Population B. Panels (c) and (d) show the distributions of outcome prevalences in subgroups determined by a risk model based on two of the four covariates (Model 1). Panels (e) and (f) show outcome prevalences in subgroups obtained by expanding Model 1 with an additional covariate (Model 2). (See Appendix for details).

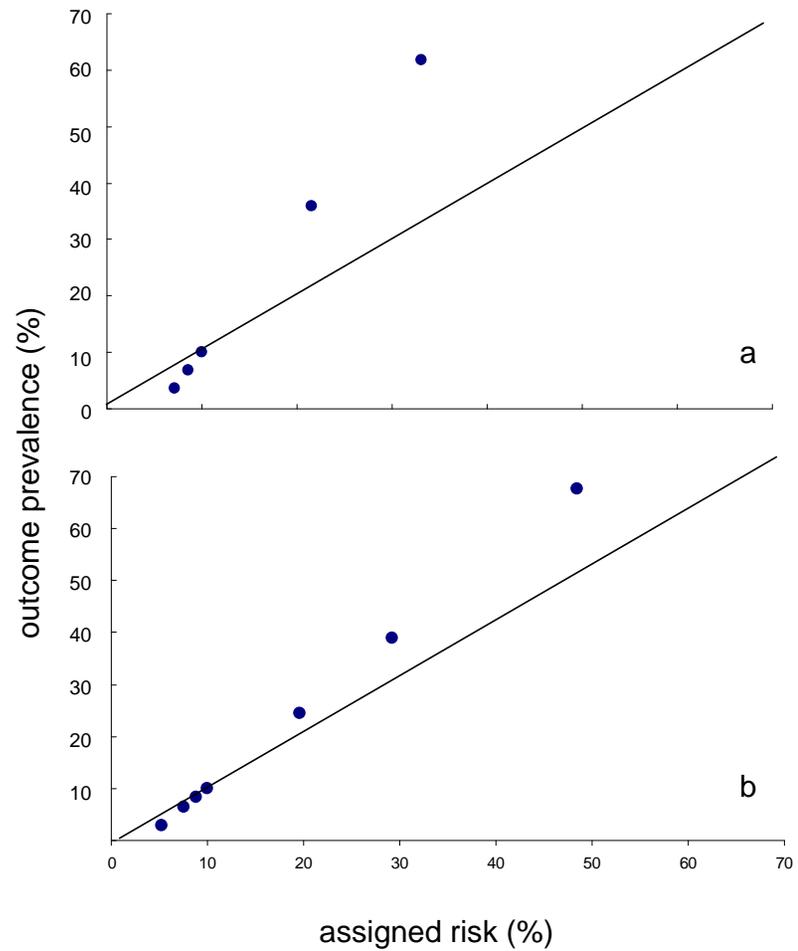

**Plate 3.** Panel (a) gives an attribute diagram for a model that uses two of the four covariates to partition Population B into five risk groups, and then assigns each group the outcome prevalence observed in Population A. Panel (b) shows the attribute diagram obtained when this model is expanded to include an additional covariate, and again is well calibrated to Population A and applied to Population B.

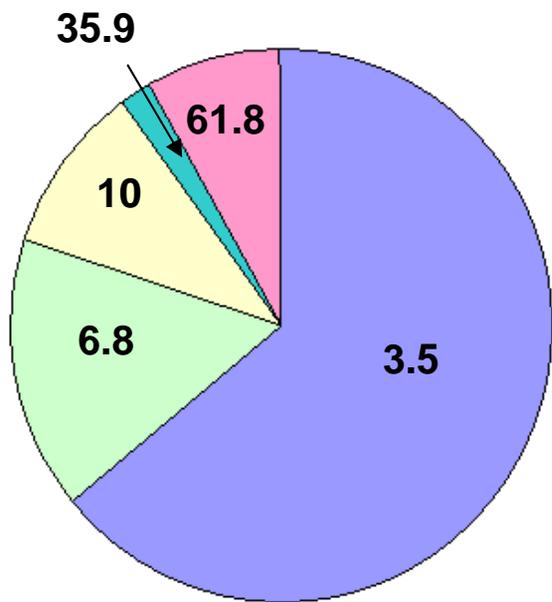
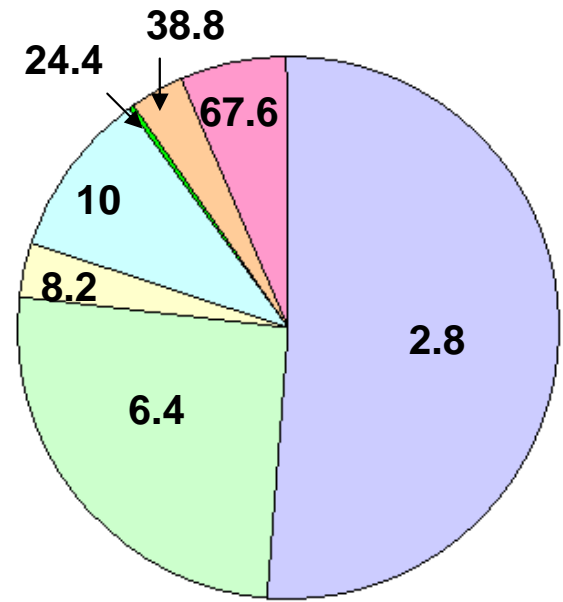
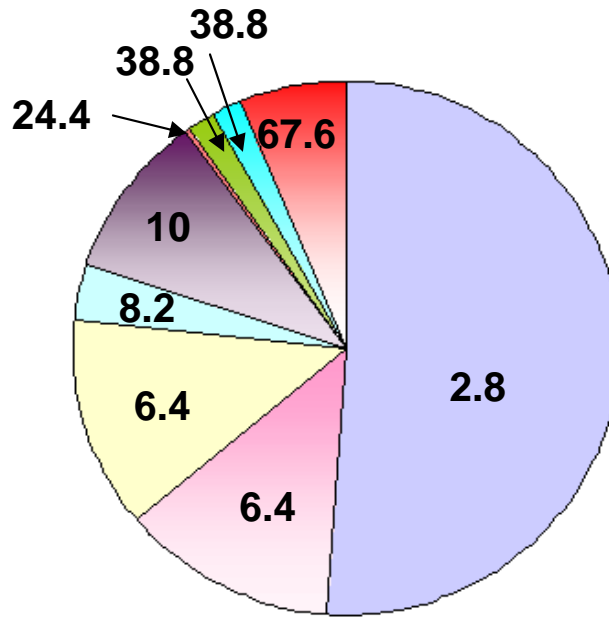

**Plate 4.** Cartoon showing how Population B is partitioned into the five risk groups of Model 1 and the seven risk groups of Model 2. Also shown is the partition induced by the cross-classified model consisting of individuals assigned a common risk by each of the two models. Risk groups are labeled with their outcome prevalences.

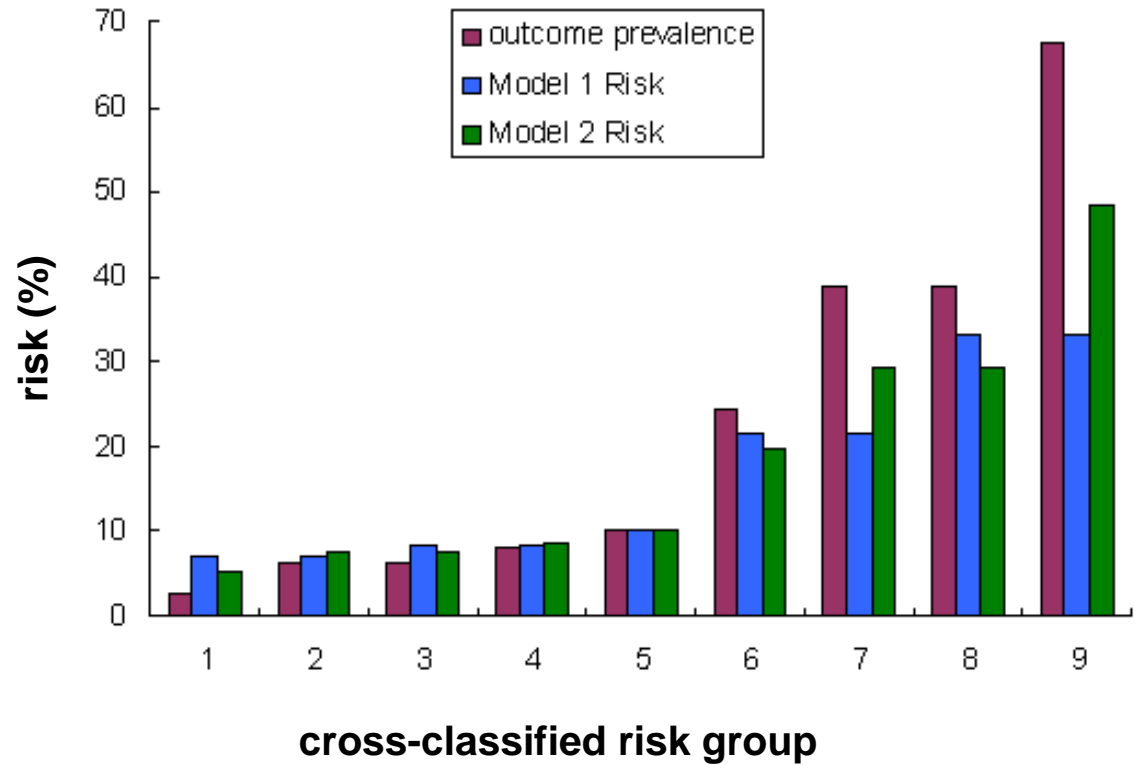

**Plate 5.** Outcome prevalences and assigned risks in subgroups determined by the cross-classified model for Models 1 and 2 when both models are calibrated to Population A and applied to Population B.

## Table 1. Precision Measures (%) for Risks of Figure 1

|  | Variance of Outcome Prevalences | Precision Loss | Risk Outcome Correlation | Concordance |
|---|---|---|---|---|
|  | Population A / Population B | Population A / Population B | Population A / Population B | Population A / Population B |
| Model 1 [a] | 0.24 / 2.57 | 8.76 / 6.43 | 16.3 / 53.4 | 59.4 / 80.6 |
| Model 2 | 0.27 / 2.70 | 8.73 / 6.30 | 17.2 / 54.7 | 59.9 / 82.1 |
| Model P [b] | 0.30 / 2.83 | 8.70 / 6.17 | 18.1 / 56.1 | 60.3 / 83.3 |

a) risks assigned using covariates $z_0, z_1$ of equation (20) (Model 1) or covariates $z_0, z_1, z_2$ of equation (21) (Model 2)
b) each individual is assigned his true risk given by equation (19)

**Table 2. Range and Standard Deviation (SD) of Cross-classified Outcome Prevalences B [a], by Risk Group of Model 1**

| Risk Group | Model 1 Outcome Prevalences (%) | Cross-classified Outcome Prevalences (%) | |
|---|---|---|---|
| | | Range | SD |
| 1 | 3.5 | 2.8-6.4 | 1.4 |
| 2 | 6.8 | 6.4-8.2 | 0.7 |
| 3 | 10.0 | 10.0-10.0 | 0 |
| 4 | 35.9 | 24.4-38.8 | 5.8 |
| 5 | 61.8 | 38.8-67.6 | 11.5 |

a) outcome prevalences in those assigned common risks by Models 1 and 2, as shown in panels (d) and (f) of Figure 1

Table 3. Ten-year Prevalences (%) of Postmenopausal ER+ Breast Cancer [a],
cross-classified by Risk Decile of Model 1 (without Estradiol) and Model 2 (with Estradiol)

| Model 1 Risk Decile | Model 2 Risk Decile | | | | | | | | | | Cross-classified Prevalences | | |
|---|---|---|---|---|---|---|---|---|---|---|---|---|---|
| | 1 | 2 | 3 | 4 | 5 | 6 | 7 | 8 | 9 | 10 | All | Range | SD |
| 1 | 0.7 | 1.0 | - | - | - | - | - | - | - | - | 0.7 | 0.7-1.0 | 0.12 |
| 2 | 0.7 | 0.8 | 2.0 | 2.9 | - | - | - | - | - | - | 1.1 | 0.7-2.9 | 0.62 |
| 3 | - | 0.5 | 1.3 | 1.7 | 2.2 | - | - | - | - | - | 1.2 | 0.5-2.2 | 0.49 |
| 4 | - | 0.4 | 1.0 | 1.3 | 2.3 | 3.6 | - | - | - | - | 1.5 | 0.4-3.6 | 0.72 |
| 5 | - | - | 1.4 | 0.7 | 1.5 | 2.7 | 6.1 | - | - | - | 1.8 | 0.7-6.1 | 1.16 |
| 6 | - | - | - | 1.0 | 1.1 | 1.5 | 3.0 | 5.2 | - | - | 1.9 | 1.0-5.2 | 9.50 |
| 7 | - | - | - | - | 0.6 | 1.2 | 2.1 | 2.6 | 6.5 | - | 2.0 | 0.6-6.5 | 8.20 |
| 8 | - | - | - | - | - | 1.6 | 1.7 | 2.4 | 4.8 | - | 2.7 | 1.6-4.8 | 1.13 |
| 9 | - | - | - | - | - | - | 2.8 | 1.5 | 3.0 | 5.7 | 3.1 | 1.5-5.7 | 1.24 |
| 10 | - | - | - | - | - | - | - | - | 1.9 | 4.8 | 4.4 | 1.9-4.8 | 1.06 |
| All | 0.7 | 0.8 | 1.4 | 1.4 | 1.6 | 1.8 | 2.4 | 2.4 | 3.3 | 5.0 | - | - | - |

a) from Rosner et al (2008)



\end{document}